\documentclass[aps,showpacs,pra,reprint,superscriptaddress]{revtex4-1}
\usepackage{stmaryrd}
\usepackage{amsfonts}
\usepackage{mathrsfs}
\usepackage{amsmath}
\usepackage{amscd}
\usepackage{graphicx}
\usepackage{booktabs}
\usepackage{subfigure}
\usepackage{natbib}

{\large {\tiny }}

\begin{document}
\title{Self-Sustained Oscillation and Dynamical Multistability of Optomechanical Systems in the Extremely-Large-Amplitude Regime}
\author{Ming Gao}
\affiliation{State Key Laboratory of Low-dimensional Quantum Physics and Department of Physics, Tsinghua University, Beijing 100084, China}
\author{Fu-Chuan Lei}
\affiliation{State Key Laboratory of Low-dimensional Quantum Physics and Department of Physics, Tsinghua University, Beijing 100084, China}
\author{Chun-Guang Du}
\affiliation{State Key Laboratory of Low-dimensional Quantum Physics and Department of Physics, Tsinghua University, Beijing 100084, China}
\author{Gui-Lu Long}
\email{gllong@mail.tsinghua.edu.cn}
\affiliation{State Key Laboratory of Low-dimensional Quantum Physics and Department of Physics, Tsinghua University, Beijing 100084, China}
\affiliation{Tsinghua National Laboratory of Information Science and Technology, Beijing 100084, China}
\affiliation{Collaborative Innovation Center of Quantum Matter, Beijing 100084, China}

\begin{abstract}

Optomechanics concerns with the coupling between optical cavities and mechanical resonators.
Most early works are concentrated in the physics of optomechanics in the small-displacement regime and consider one single optical cavity mode participating in the optomechanical coupling.
In this paper, we focus on optomechanics in the extremely-large-amplitude regime in which a mechanical resonator is coupled with multiple optical cavity modes during the oscillation.
We explicitly show that the mechanical resonator can present self-sustained oscillations in a novel way with limit cycles in the shape of sawtooth-edged ellipses and  exhibit dynamical multistability.
By analyzing the  mechanical oscillation process and the accompanied variation of the optical cavity occupation, we develop an energy-balanced condition to ensure the stability of self-sustained oscillation. The effect of the mechanical nonlinearities on the dynamics of the mechanical resonator is also investigated.

\end{abstract}
\maketitle

\section{INTRODUCTION}

Optomechanics has attracted much attention and is undergoing rapid development in the recent years \cite{review-oe,review-science,review-physics,review-npho,review-introduction-quantum,review-phyrep,review-physics-today,review-arxiv}. It concerns with the coupling between optical cavity modes  and mechanical degrees of freedom via radiation pressure \cite{radiation-prl-1999}, optical gradient forces \cite{gradient-force}, photothermal forces \cite{photothermal-nature-2004} or the Doppler effect \cite{doppler-prl-2008}. Various motivations have driven the development of this research field, such as detection of gravitational wave \cite{gravitational-pla-2001,gravitational-pla-2002,gravitational-pla-2002-2,gravitational-pra-2010}, more sensitive sensors of displacement, mass or force \cite{sensor-prl-2006,sensor-pra-2006,sensor-prl-2010,sensor-pra-2010,scpmaresponse,lsawater}, fundamental studies of quantum mechanics \cite{testquantum-epl-2001,testquantum-prl-2007,scpmaeit}, preparation of macroscopic quantum state \cite{quantum-state-prl-2002,testquantum-prl-2003,aplhwy3wave,praliuyx}, quantum state transfer \cite{transfer-pra-2010,transfer-prl-2012-e,transfer-prl-2012,transfer-ncom-2012,transfer-science-2012}, novel nonlinear coupling quantum physics \cite{single-photon-prl-2011,photon-blockade-prl-2011,lsameta}, and quantum information processing \cite{information-prl-2010}.
Most of these motivations focus on the quantum level of optomechanical systems and require cooling the mechanical resonator as close as possible to its ground state, which has been widely investigated both theoretically \cite{cooling-prl-2007-1,cooling-prl-2007-2} and experimentally \cite{cantilever-nature-e,cantilever-nature-1,cantilever-nature-2,radiation-prl-2006,cooling-nphy-2009-1,cooling-nphy-2009-2,cooling-nphy-2009-3,cooling-nature-2012}.

On the other hand, amplification of the mechanical oscillation is very useful for both practical applications and fundamental researches, such as non-volatile mechanical memory \cite{memory-nnano-2011}, synchronization of remote mechanical resonators \cite{synchronization-prl-2012} and chaos dynamics \cite{chaos-prl-2005,chaos-prl-2007}. With a driving laser of frequency blue detuned with respect to the cavity resonance and power above a certain threshold, the mechanical resonator can run into self-sustained oscillations \cite{chaos-prl-2005,sso-prl-2005,sso-oe-2005}.
An arbitrary tiny thermal fluctuation will be amplified into an oscillation with exponentially increasing amplitude, and finally be saturated into a stable periodic oscillation. Self-sustained oscillation is of broad interests. It not only exists in optomechanical system, but also in other systems such as a resonator driven by a superconducting single-electron transistor \cite{sso-superconducting-prl-2007}, an ultracold atomic gas in an optical cavity \cite{sso-coldatom-nphy-2008}, or a metallic point contact deposited on a metallic spin-valve stack \cite{sso-metallic-prl-2008}.
When optomechanical systems present self-sustained oscillations, dynamical multistability may emerge as theoretically predicted in \cite{multistability-prl-2006} and experimentally explored in \cite{multistability-prl-2008}, which means that there may exist multiple possible stable oscillations at a set of fixed parameters.

Most early theoretical works on optomechanical self-sustained oscillation focus attention on the small-displacement regime.
In this regime, it is sufficient to take only one single optical cavity mode into account participating in the coupling with a mechanical resonator, and the dependence of the resonance frequency of this mode on the displacement of the mechanical resonator can be treated  linearly \cite{multistability-prl-2006}.
The mechanical resonator conducts approximately sinusoidal oscillations at its intrinsic frequency, and therefore, its limit cycles in the phase space are approximately elliptical.
While in the large-amplitude regime, which can be realized in an optomechanical system driven by a high-power laser, the mechanical resonator can display different self-sustained oscillations with limit cycles mushroom-like in shape \cite{large-amplitude-pra-2012}.

If the power of the driving laser is further increased, the amplitude of the mechanical oscillation can be comparable with the wavelength of the laser, i.e., the system reaches the extremely-large-amplitude regime (ELAR).
Multiple optical cavity modes of different orders may be excited and participate the coupling with the mechanical resonator during the oscillation.
In this work, we will focus on optomechanics in the ELAR.
We organize the paper in the following way.
Sec. \ref{sec2} introduces the model and the Hamiltonian, and gives the dynamical equations of the system.
In Sec. \ref{sec3}, the self-sustained oscillation and dynamical multistability in the ELAR  are studied by analyzing the limit cycles of the mechanical resonator in the phase space.
An energy-balanced condition is given to ensure the stability of self-sustained oscillation.
The effect of the mechanical nonlinearities on the dynamics of the mechanical resonator is discussed in Sec. \ref{sec4}.
Finally, Sec. \ref{sec5} gives a summary of this work.

\section{Hamiltonian and Dynamical Equations}
\label{sec2}

\begin{figure}[!ht]
\centering
\includegraphics[width=2.9 in,height=1.9 in]{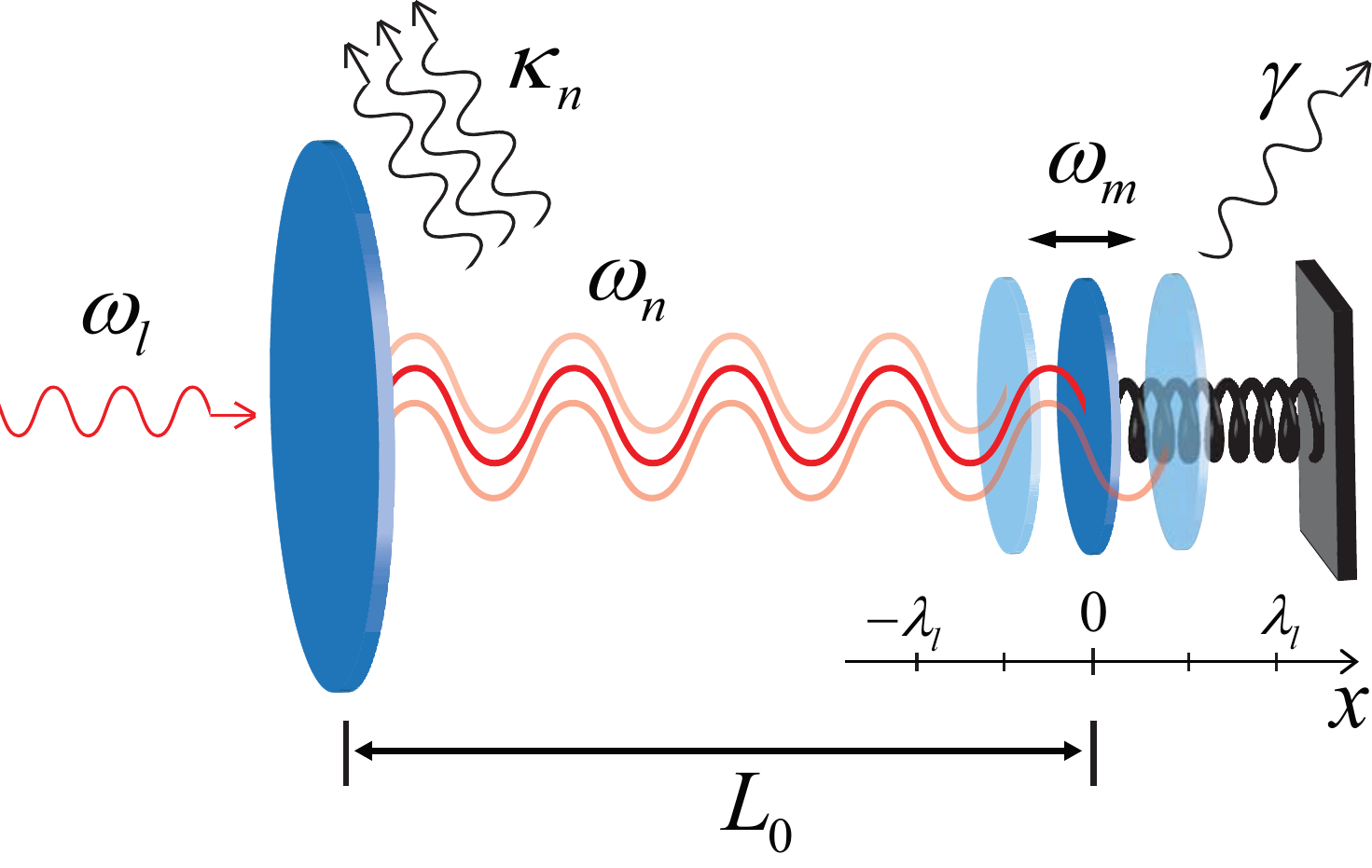}
\caption{(Color online) Schematic of a generic optomechanical system. An optical cavity of static equilibrium length $L_0$ with a fixed, partially reflecting mirror on the left side and a moveable, perfectly reflecting mirror on the right side is driven by an external laser of frequency $\omega_l$ and power $P$.
In the ELAR, multiple optical cavity modes of frequencies $\omega_n$ and decay rates $\kappa_n$ can be excited and coupled with the mechanical resonator of intrinsic frequency $\omega_m$, mass $m$ and damping rate $\gamma$. }
\label{fig1}
\end{figure}

We consider a generic optomechanical system as shown in Fig. \ref{fig1}, which is essentially a Fabry-P\'{e}rot cavity with a fixed, partially reflecting mirror on one side and a movable, perfectly reflecting mirror on the other side.
In this system, radiation pressure provides the dominant optomechanical coupling which is typically dispersive, implying that the primary effect of the moveable mirror is to shift the frequency of the optical cavity modes.
The cavity is driven by an external laser of frequency $\omega_l$ and power $P$.
If the driving laser is turned off, the static equilibrium length of the cavity is $L_0$.
The movable mirror in this model can be considered as a mechanical resonator of intrinsic frequency $\omega_m$, mass $m$ and damping rate $\gamma$.
Usually, the displacement $x$ of the mechanical resonator is assumed to be very small, so that it is sufficient to take only one single optical catity mode of frequency $\omega_c(x)$ and decay rate $\kappa$ into account in the coupling with the mechanical resonator.
In this small-displacement regime, $\omega_c(x)$ is approximately equal to the first-order expansion around the static equilibrium position $x=0$,
\begin{equation}
\omega_c(x)\approx\omega_c(0)+\omega_c^{\prime}(0)x=\omega_{c0}-(\omega_{c0}/L_0) x,\nonumber
\end{equation}
where $\omega_{c0}=\omega_c(0)$.
In the frame rotating at frequency $\omega_l$, the Hamiltonian of the system can be written as
\begin{eqnarray}
\label{eq1}
H&=&\hbar(\omega_{c0}-\omega_l)\hat{a}^\dag \hat{a}+\frac{\hat{p}^2}{2m}+\frac{1}{2}m\omega_{m}^2 \hat{x}^2 -\hbar g \hat{a}^\dag \hat{a} \hat{x} \nonumber \\
&&+\hbar\alpha_L(\hat{a}+\hat{a}^\dag)+H_{\kappa}+H_{\gamma} ,
\end{eqnarray}
where $\hat{a}$ and $\hat{a}^\dag$ are the bosonic annihilation and creation operators of the optical cavity mode, $\hat{x}$ and $\hat{p}$ are the position and momentum operators of the mechanical resonator, $g=\omega_{c0}/L_0$ is the optomechanical coupling, $\alpha_L$ is the complex amplitude of the driving laser field which satisfies $|\alpha_L|^2=2\kappa P/\hbar \omega_c$.
Here, without loss of generality, $\alpha_L$ is set to be real.
$H_{\kappa}$ denotes the coupling between the optical cavity mode and the vacuum bath that leads to the decay rate $\kappa$.
$H_{\gamma}$ refers to the interaction between the mechanical resonator and the thermal reservoir which is the cause of the damping rate $\gamma$.

If the optical cavity is driven by a high-power laser, the small-displacement assumption will be no longer valid and the system will reach the large-amplitude regime. The expansion of $\omega_c(x)$ to the first-order will not be a good approximation.
It is necessary to deal with $\omega_c(x)$ directly without any approximate expansion.
If the power $P$ is high enough, the amplitude $A$ of the mechanical oscillation may be comparable with the wavelength $\lambda_l$ of the driving laser,
\begin{equation}
A/\lambda_l\sim 1,\nonumber
\end{equation}
i.e., the system reaches the ELAR.
In this case, not only  $\omega_c(x)$ should be dealt with directly,  but also multiple optical cavity modes of different orders should be taken into account in the coupling with the mechanical resonator during the oscillation.
Concretely, each time the mechanical resonator passes through the positions that satisfy $x+L_0=n\lambda_l/2$ ($n\in\mathbb{N}$), the $n$th-order optical cavity mode of frequency $\omega_n(x)=n\pi c/(x+L_0)$ will be excited.
So during a whole cycle of the mechanical oscillation, a series of optical cavity modes will be excited.
All these modes are coupled with the mechanical resonator by radiation pressure.
Here, we assume that the size of the mechanical resonator is much larger than the amplitude of the oscillation.
In this situation, we can treat the mechanical resonator as a harmonic resonator and neglect the effect of mechanical nonlinearities (which will be considered in Sec. \ref{sec4}).
Thus in the ELAR, the Hamiltonian of the system reads,
\begin{eqnarray}
\label{eq2}
H&=&\hbar\sum_{n=1}^{\infty}[\omega_n(x)-\omega_l]\hat{a}_n^\dag \hat{a}_n+\frac{\hat{p}^2}{2m}+\frac{1}{2}m\omega_{m}^2 \hat{x}^2\nonumber\\
&&+\hbar\alpha_L\sum_{n=1}^{\infty}(\hat{a}_n+\hat{a}_n^\dag)+\sum_{n=1}^{\infty}H_{\kappa_n}+H_{\gamma} ,
\end{eqnarray}
where $\hat{a}_n$ and $\hat{a}_n^\dag$ are the bosonic annihilation and creation operators of the $n$th-order optical cavity mode of frequency $\omega_n(x)$ and decay rates $\kappa_{n}$.
Heisenberg equations of motion for operators $\hat{a}_n$, $\hat{x}$ and $\hat{p}$ can be easily derived from the Hamiltonian above.
In this paper, we aim to investigate the purely classical dynamics of the system, so we replace $\hat{a}_n$ by the complex coherent light amplitude $\alpha_n(t)$, as well, $\hat{x}$ and $\hat{p}$ by their classical counterpart $x(t)$ and $p(t)$.
We thus obtained the classical dynamical equations of the system as follows,
\begin{eqnarray}
\label{eq3}
\dot{\alpha}_n&=&-i\left(\frac{n\pi c}{x+L_0}-\omega_l\right)\alpha_n-i\alpha_L-\kappa_n\alpha_n , \\
\label{eq4}
\ddot{x}&=&-\omega_m^2x+\frac{\hbar}{m}\sum_{n=1}^{\infty}\frac{n\pi c}{(x+L_0)^2}|\alpha_n|^2-\gamma\dot{x} .
\end{eqnarray}
For simplification, in the following, we assume all the $\kappa_n$ are equal, namely,  $\kappa_n=\kappa$.

\section{Self-Sustained oscillation and Dynamical Multistability}
\label{sec3}

Eqs. (\ref{eq3}) and (\ref{eq4}) are two coupled nonlinear differential equations, it is difficult to derive analytical solutions for them.
Therefore we integrate these two equations numerically by using a fourth-order Runge-Kutta algorithm.
We have considered the experimental feasibility \cite{experiment-apl-2007,experiment-nature-2008,experiment-nature-2011,experiment-apl-2011} and set the parameters as follows, the intrinsic frequency, mass, and damping rate of the mechanical resonator are $\omega_m=10^7$ Hz, $m=5\times10^{-15}$ kg, and $\gamma=10^{-2}\omega_m$.
The wavelength and frequency of the external driving laser are $\lambda_l=1000$ nm and $\omega_l=2\pi c/\lambda_l$. The static equilibrium length of the cavity is $L_0=N\lambda_l/2$ and $N=10000$.
The decay rates of the cavity modes are $\kappa_n=\kappa=10^2\omega_m$, i.e., the system is in the unresolved sideband regime so that the optical cavity can respond quickly enough to the fast mechanical oscillation.
It should be noticed here that, compared with $L_0$, the displacement of the moving mirror (i.e., the change of the cavity length) is usually very small even in the ELAR. Thus, we treat $\kappa$ as a constant rather than a function of $x$.

\begin{figure}[!htp]
\centering
\includegraphics[width=3.4 in]{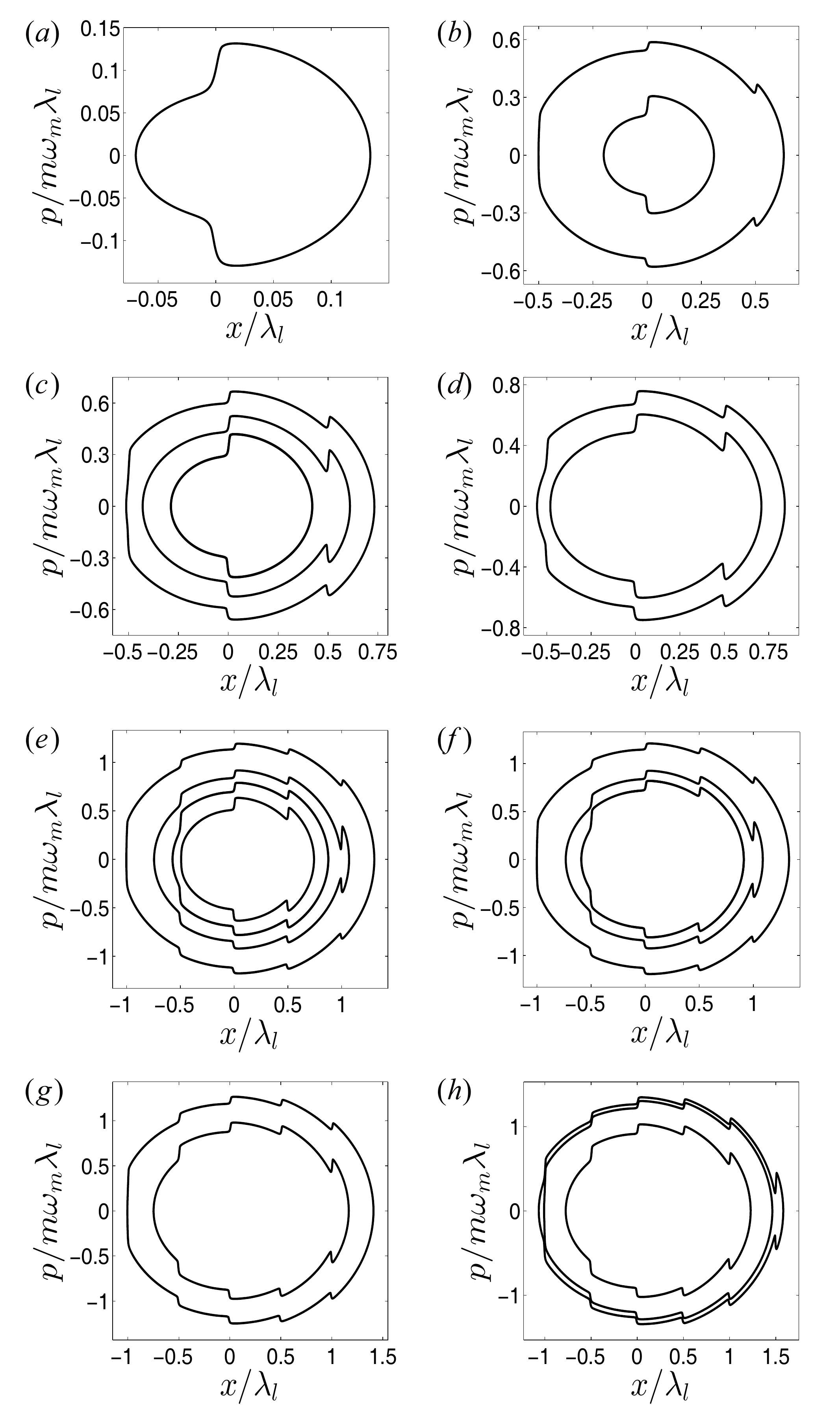}
\caption{Limit cycles in the phase space of the mechanical resonator scanned by $x$ and $p=m\dot{x}$ with different values of $P$.
The parameters are: $\omega_m=10^7$ Hz, $m=5\times10^{-15}$ kg, $\gamma=10^{-2}\omega_m$, $\kappa_n=\kappa=10^2\omega_m$, $\lambda_l=1000$ nm, $L_0=N\lambda_l/2$ and $N=10000$. The values of $P$ are: (a)$P=1$ W, (b)$P=4.1$ W, (c)$P=7$ W, (d)$P=10$ W, (e)$P=11$ W, (f)$P=12$ W, (g)$P=15$ W, (h)$P=17$ W, as specified by the vertical black dashed lines in Fig. \ref{fig4}.
}
\label{fig2}
\end{figure}

From the numerical solutions of Eqs. (\ref{eq3})-(\ref{eq4}), we plot the limit cycles in the phase space of the mechanical resonator scanned by $x$ and $p=m\dot{x}$ as shown in Fig. \ref{fig2}.
Each limit cycle corresponds to a stable self-sustained oscillation in the long-time limit.
It is shown that the mechanical resonator can present self-sustained oscillations in a novel way with limit cycles in the shape of sawtooth-edged ellipses.
When $P$ is relatively low, there is only one dynamical stable solution of the coupled Eqs. (\ref{eq3})-(\ref{eq4}) in the long-time limit, hence there is only one limit cycle in the phase space of the mechanical resonator as shown in Fig. \ref{fig2}(a).
The limit cycle is mushroom-like. Self-sustained oscillations with limit cycles of this shape have been theoretically studied \cite{large-amplitude-pra-2012} and experimentally observed \cite{ombec-science-2008} in optomechanical systems.

\begin{figure}[!htp]
\centering
\includegraphics[width=3.2 in]{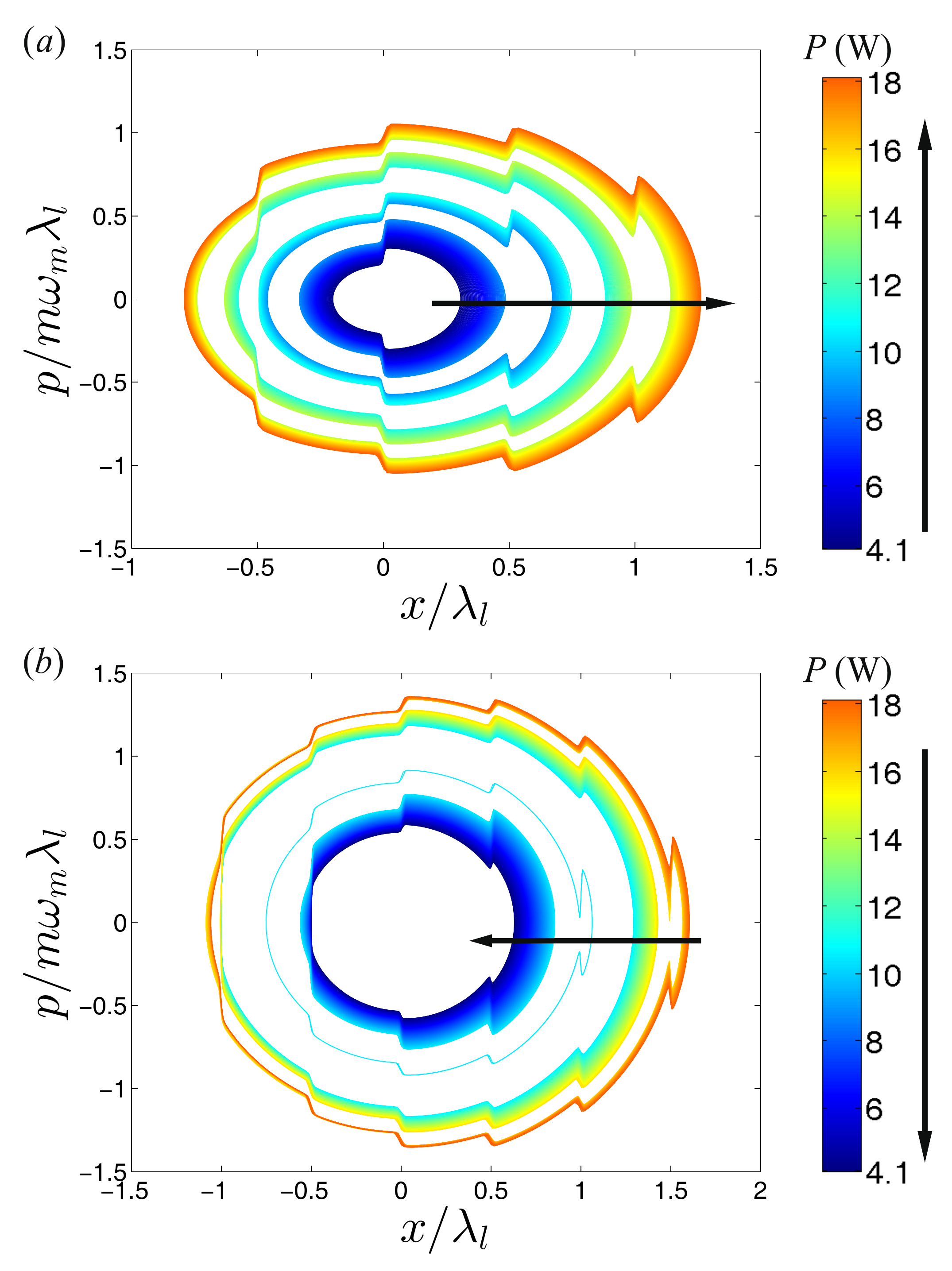}
\caption{(Color online) The variation of the minimum(a) and maximum(b) limit cycles as the power of the external driving power changes in the range of 4.1W to 18W in which the mechanical resonator presents dynamical multistability.
}
\label{fig3}
\end{figure}

When $P$ is above 4.1W, there are multiple different limit cycles in the phase space of the mechanical resonator as shown in Figs. \ref{fig2}(b)-(h), meaning that the mechanical resonator exhibits dynamical multistability.
In Fig. \ref{fig3}, we show the variation of the minimum and maximum limit cycles respectively as $P$ changes in the range from 4.1W to 18W.
If the mechanical resonator initially reach a limit cycle in Fig. \ref{fig3}(a) and $P$ is then increased slowly, the mechanical resonator will oscillate with an expanding limit cycle changing shape in the way shown in Fig. \ref{fig3}(a).
While, if the mechanical resonator is initially trapped in a limit cycle in Fig. \ref{fig3}(b), it will oscillate with an shrinking limit cycle as $P$ decreased gradually as indicated in Fig. \ref{fig3}(b).

\begin{figure}[!htp]
\centering
\includegraphics[width=3.4 in]{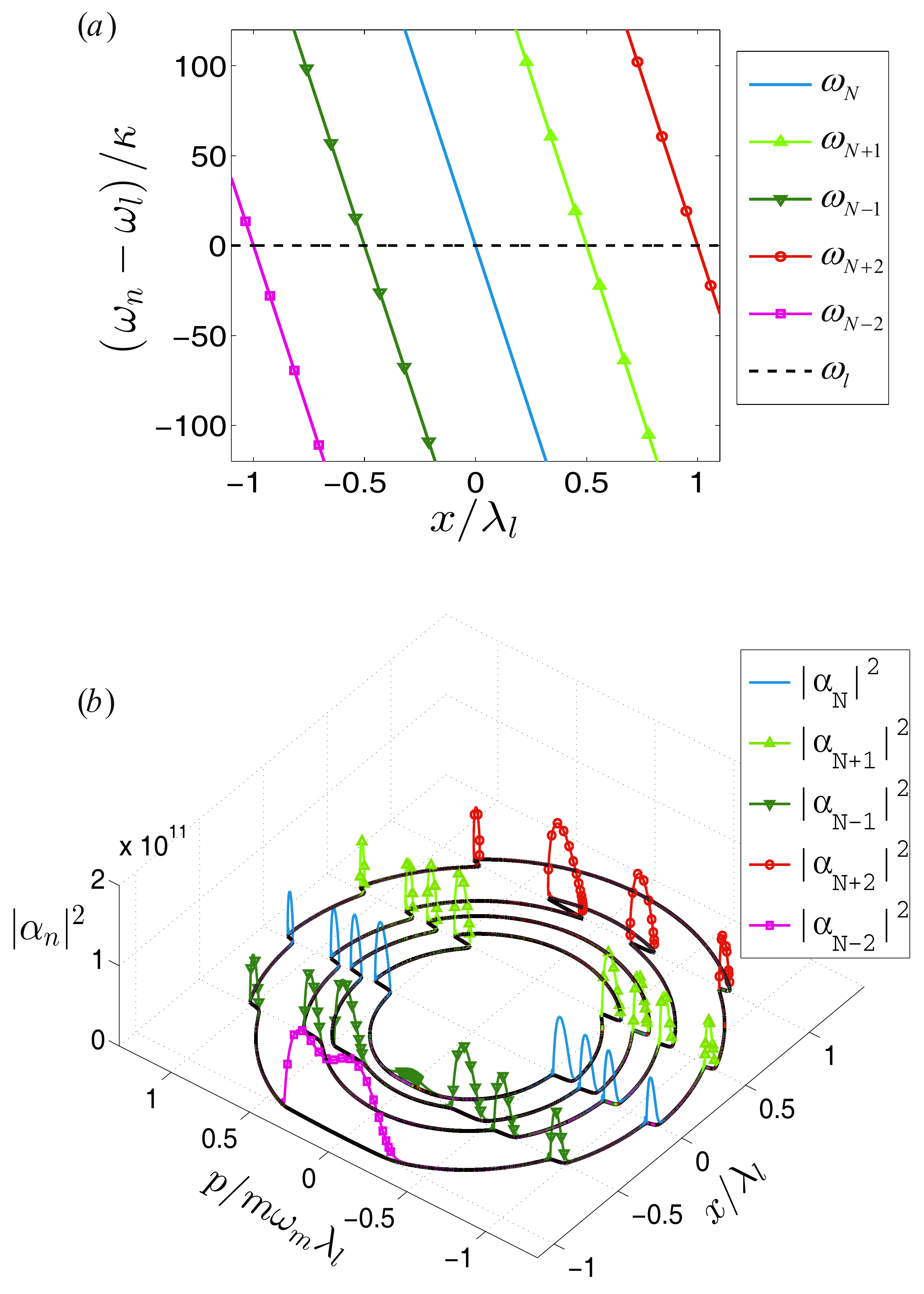}
\caption{(Color online) (a)The dependence of the frequencies $\omega_n$ of the optical cavity modes on the displacement $x$ of the mechanical resonator. (b)The variation of the optical cavity occupation with the oscillation of the mechanical resonator. Here, the power of the external driving laser is $P=11$ W. The other parameters are same as in Fig. \ref{fig2}.
}
\label{fig4}
\end{figure}

When the mechanical resonator moves forward (to the right as shown in Fig. \ref{fig1}), the frequencies of the optical cavity modes $\omega_n(x)=n\pi c/(x+L_0)=n\pi c/(x+N\lambda_l/2)$ decrease.
Conversely, when it moves backward, the frequencies increase.
Every time the mechanical resonator passes through the positions $x=x_k=k\lambda_l/2$ ($k=...,-2,-1,0,1,2,...$), the frequency of the optical cavity mode of order $N+k$ meets the external driving laser frequency $\omega_l$ as shown in Fig. \ref{fig4}(a).
As a result, this optical cavity mode is excited and the photon number in the cavity becomes very large rapidly, accordingly, the radiation pressure exerted by light stored in the cavity increases rapidly.
When the mechanical resonator goes away from these positions, the cavity occupation as well as the radiation pressure decreases rapidly to almost zero due to the large decay rates of the cavity modes.
The variation of the photon number in the cavity accompanied with the oscillation of the mechanical resonator is shown in Fig. \ref{fig4}(b).

When the mechanical resonator is away from positions $x=x_k$, it approximately carries out damped harmonic oscillation,
\begin{eqnarray}
\label{eq5}
\ddot{x}&=&-\omega_m^2x-\gamma\dot{x} .
\end{eqnarray}
Every time the mechanical resonator passes through positions $x=x_k$, radiation pressure in the cavity kicks it and dynamical equations of the system can be approximately written as,
\begin{eqnarray}
\label{eq6}
\dot{\alpha}_{_{N+k}}&=&i g_{_{N+k}}\left(x-x_k\right)\alpha_{_{N+k}}-i\alpha_L-\kappa\alpha_{_{N+k}} , \\
\label{eq7}
\ddot{x}&=&-\omega_m^2x+\frac{\hbar g_{_{N+k}}}{m}|\alpha_{_{N+k}}|^2-\gamma\dot{x} ,
\end{eqnarray}
where $g_{_{N+k}}=4\pi c/[(N+k)\lambda^2]$ is the optomechanical coupling strength.
It can be derived from Eq. (\ref{eq7}) that,
\begin{eqnarray}
\label{eq8}
|\alpha_{_{N+k}}|^2 &\approx &\frac{\alpha_L^2}{g_{_{N+k}}^2\left(x-x_k\right)^2+\kappa^2}
\nonumber\\
&&\times\left[1+\frac{4\kappa g_{_{N+k}}^2\left(x-x_k\right)\dot{x}}{\left[g_{_{N+k}}^2\left(x-x_k\right)^2+\kappa^2\right]^2}\right] .
\end{eqnarray}
\begin{figure}[!htp]
\centering
\includegraphics[width=3.2 in]{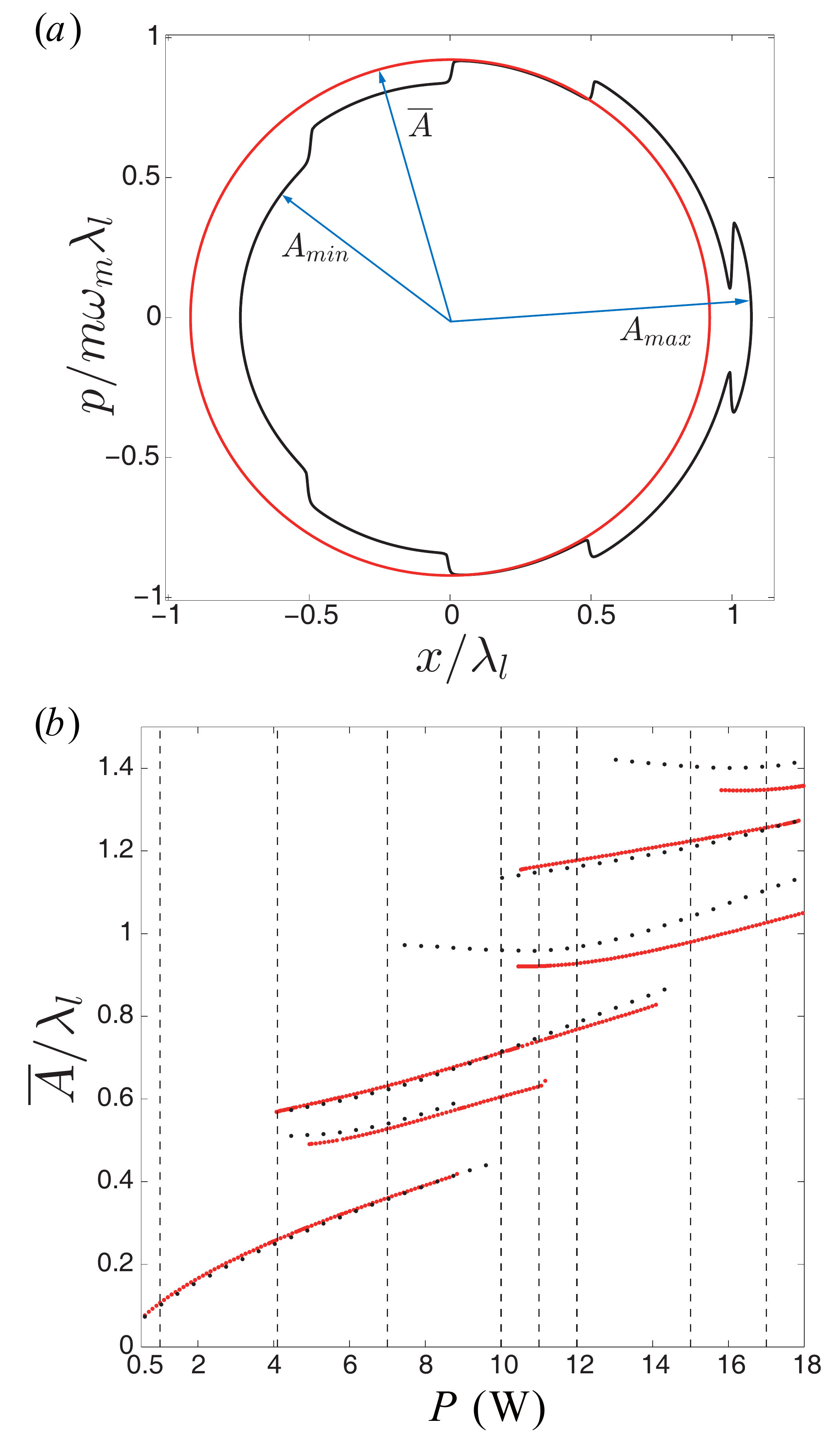}
\caption{(Color online) (a) Definition of average amplitude as $\overline{A}=\sqrt{(A_{min}^2+A_{max}^2)/2}$.
(b) Attractor diagram on a plane of $\overline{A}$ and $P$.
Here, the parameters are same as in Fig. \ref{fig2}.
The red (dense) dots show the numerically exact results captured by solving the original coupled dynamical equations (\ref{eq3})-(\ref{eq4}).
The black (sparse) dots are the approximately estimated results (see the Appendix) from the energy-balanced condition, Eq. (\ref{eq12}).
The vertical black dashed lines specify some values of $P$ distributed in different regions.
Limit cycles for these values of $P$ are plotted in Fig. \ref{fig2}.
}
\label{fig5}
\end{figure}
Thus, the mechanical resonator satisfies the following equation of motion,
\begin{eqnarray}
\label{eq9}
\ddot{x} &=&-\omega_m^2x-\gamma\dot{x}+\frac{\hbar g_{_{N+k}}\alpha_L^2}{m\left[g_{_{N+k}}^2\left(x-x_k\right)^2+\kappa^2\right]}
\nonumber\\
&&\times\left[1+\frac{4\kappa g_{_{N+k}}^2\left(x-k\lambda /2\right)\dot{x}}{\left[g_{_{N+k}}^2\left(x-x_k\right)^2+\kappa^2\right]^2}\right] .
\end{eqnarray}
If the mechanical resonator moves forward, radiation pressure do positive work on it:
\begin{eqnarray}
\label{eq10}
W_{_{N+k}}^{+} &=& \frac{\hbar\alpha_L^2\pi}{\kappa}+\frac{3\hbar^2\alpha_L^4 g_{_{N+k}}\pi}{8m\kappa^4\dot{x}_{k}^-} .
\end{eqnarray}
where $\dot{x}_{k}^-$ is the velocity of the mechanical resonator just before it passes forward through the position $x=x_k$.
As a result, the kicks lead to sharp accelerations of the mechanical resonator.
On the contrary, if the mechanical resonator moves backward, radiation pressure do negative work on it,
\begin{eqnarray}
\label{eq11}
W_{_{N+k}}^{-} &=& -\frac{\hbar\alpha_L^2\pi}{\kappa}-\frac{3\hbar^2\alpha_L^4 g_{_{N+k}}\pi}{8m\kappa^4\dot{x}_{k}^+} ,
\end{eqnarray}
where $\dot{x}_{k}^+$ is the velocity of the mechanical resonator just before it passes backward through  position $x=x_k$.
So the mechanical resonator experiences sharp decelerations.
As a reflection of these processes, there are some sawteeth on the limit cycles at  positions $x=x_k$ as shown in Fig. \ref{fig2}.
If $x_k$ is out of the oscillation range of the mechanical resonator, we have $W_{_{N+k}}^{+}=W_{_{N+k}}^{-}=0$.
When the total net work did by radiation pressure is balanced with the dissipative energy $E_\gamma$ during one whole cycle:
\begin{eqnarray}
\label{eq12}
\sum_{k}\left(W_{_{N+k}}^{+}+W_{_{N+k}}^{-}\right) &=& E_\gamma ,
\end{eqnarray}
the mechanical resonator reaches stable self-sustained oscillation.
At some parameters, there may simultaneously exist multiple different possible stable oscillations that can satisfy this energy-balance condition.
So the mechanical resonator may exhibit dynamical multistability.

To demonstrate dynamical multistability concisely, we define average amplitude as $\overline{A}=\sqrt{(A_{min}^2+A_{max}^2)/2}$ as shown in Fig. \ref{fig5}(a), where $A_{min}$($A_{max}$) is the minimum(maximum) amplitude of a limit cycle.
It should be noticed that with the effect of radiation pressure the dynamical equilibrium position $\overline{x}$ is shifted from the static equilibrium position $x=0$, but in the ELAR, the shift is small and can be neglected compared with the amplitude.

Fig. \ref{fig5}(b) shows the dependence of $\overline{A}$ on $P$, which can be considered as an attractor diagram.
The red (dense) dots refer to the numerically exact results captured by solving the original coupled dynamical equations (\ref{eq3})-(\ref{eq4}).
The black (sparse) dots denote the approximately estimated results (for details, see the Appendix) from the energy-balanced condition, Eq. (\ref{eq12}).
The discrete average amplitudes of the mechanical resonator reveals that the energy-balance condition leads to an amplitude locking effect.
When $P$ is above 4.1W, the mechanical resonator exhibits dynamical multistability, so $\overline{A}$ can take multiple values for
a fixed $P$ as shown in Fig. \ref{fig5}(b).

\section{Mechanical Nonlinearities}
\label{sec4}

\begin{figure}[!htp]
\centering
\includegraphics[width=3.2 in]{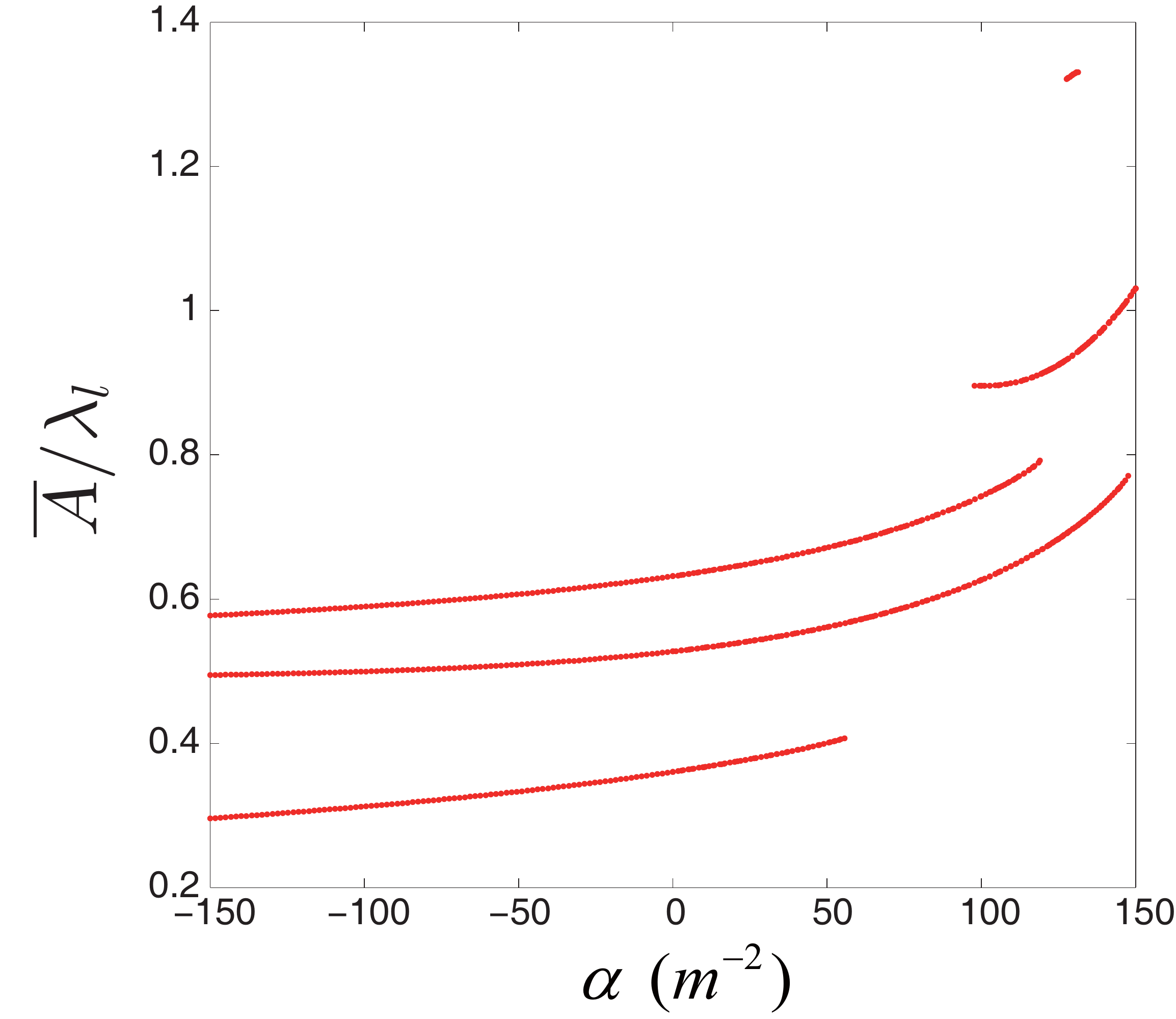}
\caption{(Color online) The dependence of the average amplitude $\overline{A}$ on $\alpha$.
Here, the power of the external driving laser is set to be $P=7$ W.
The other parameters are same as in Fig. \ref{fig2}.
}
\label{fig6}
\end{figure}
In ELAR, if the size of the mechanical resonator is not very large, the stress induced in it during the oscillation may be strong and the resulting mechanical nonlinearities \cite{nonlinear-njp-2012,nonlinear-prl-2013,nonlinear-pra-2014} may be non-neglectable.
In this situation, the dynamics of the mechanical resonator can be described by the following Duffing equation, including optomechanical coupling,
\begin{eqnarray}
\label{eq13}
\ddot{x}&=&-\omega_m^2(1+\alpha x^2)x+\frac{\hbar}{m}\sum_{n=1}^{\infty}\frac{n\pi c}{(x+L_0)^2}|\alpha_n|^2\nonumber\\&&-\gamma\dot{x} ,
\end{eqnarray}
where $\alpha$ is the cubic nonlinear constant and has the units of $m^{-2}$.
If the spring constant of the mechanical resonator weakens with increasing amplitude, the value of $\alpha$ is negative.
On the contrary, a positive value of $\alpha$ indicates that the spring constant stiffens with increasing amplitude.
Fig. (\ref{fig6}) shows the dependence of the average amplitude $\overline{A}$ on $\alpha$.
When $\alpha<0$, the average amplitude $\overline{A}$ changes gently with $\alpha$, meaning that for a weakening mechanical spring, the mechanical nonlinearities have a weak effect on the mechanical oscillation.
For a stiffening mechanical spring ($\alpha>0$), the effect is also very weak when $\alpha$ is small.
Only when $\alpha$ is large, the effect becomes strong.
As $\alpha$ increases, $\overline{A}$ increases sharply and the distribution of the dynamical multistability is affected.

\section{Summary}
\label{sec5}

We have studied classical dynamics of a generic optomechanical system in the ELAR in which a mechanical resonator is coupled with multiple optical cavity modes.
It has been shown that the mechanical resonator can present self-sustained oscillations in a wide range of parametric space.
Instead of sinusoidal oscillations in the small-displacement regime, the mechanical resonator display more complicated oscillations in the ELAR with limit cycles in the shape of sawtooth-edged ellipses.
We have analyzed the process of the mechanical oscillation and the accompanied variation of the optical cavity occupation.
Based on these, we derived an energy-balanced condition of stable self-sustained oscillation.
We have demonstrated that the mechanical resonator may exhibit dynamical multistability, which can be explained by the fact that there may simultaneously exist multiple different stable oscillations that can satisfy the energy-balanced condition.
The effect of the mechanical nonlinearities on the dynamics of the mechanical resonator has also been discussed, and it has been shown that the effect is weak in a wide range of values of the nonlinearity parameter $\alpha$. Only when $\alpha$ is positive and very large, the effect becomes strong and may has an influence on the distribution of the dynamical multistability.

Dynamical multistability may find applications in sensitive force or displacement detections and memory storage. In the parametric space of our calculation in this paper, the required power of the external driving laser is very high. However, with smaller mass, intrinsic frequency or damping rate, the mechanical resonator can reach the ELAR driven by a much weaker laser field. It is possible to observe self-sustained oscillation demonstrated in this paper within the reach of current experimental technology.

\section*{ACKNOWLEDGMENTS}
We thank M. Poot for helpful discussions.
This work is supported by the National Natural Science
Foundation of China under Grant No.11474181, the National Basic Research Program of China under Grant No.
2011CB9216002.

\section*{APPENDIX}
The energy-balanced condition, Eq. (\ref{eq12}) can be rewritten as an equation of two unknown quantities $A_{min}$ and $A_{max}$ as follows:
\begin{eqnarray}
\label{eq14}
\sum_{\substack{k\\(-A_{min}\leq x_k\leq A_{max})}}\left(W_{_{N+k}}^{+}+W_{_{N+k}}^{-}\right) \nonumber\\= E_\gamma(A_{min},A_{max}) ,
\end{eqnarray}
where $W_{_{N+k}}^{+}$ and $W_{_{N+k}}^{-}$ are expressed by Eq. (\ref{eq10}) and (\ref{eq11}).
It should be noticed that $A_{min}$ and $A_{max}$ are not independent of each other, but are correlated.
Concretely, if the mechanical resonator moves forward with initial position and velocity $(x,\dot{x})=(-A_{min},0)$, it approximately experiences the following motion processes.
It carries out damped harmonic motion obeying Eq. (\ref{eq5}) until it reach a state $(x_k,\dot{x}_{k}^-)$.
Next, it experiences a sharp acceleration rapidly and its state undergoes an abrupt change:
\begin{eqnarray}
\label{eq15}
&(x_k,\dot{x}_{k}^-)\rightarrow \nonumber\\
&\left(x_k,\sqrt{(\dot{x}_{k}^-)^2+\frac{2\hbar\alpha_L^2\pi}{m\kappa}+\frac{3\hbar^2\alpha_L^4 g_{N+k}\pi}{4m^2\kappa^4\dot{x}_k^-}}\right),
\end{eqnarray}
which can be derived from Eq. (\ref{eq10}).
Then the mechanical resonator alternately experiences damped harmonic motions and sharp accelerations at the positions $x=x_k$ until it reach a unique and determinate state ($A_{max}$,0).
Similarly, when the mechanical resonator moves backward, it alternately experiences damped harmonic motions and the following sharp decelerations at the positions $x_k$:
\begin{eqnarray}
\label{eq16}
&(x_k,\dot{x}_{k}^+)\rightarrow \nonumber\\
&\left(x_k,-\sqrt{(\dot{x}_{k}^+)^2-\frac{2\hbar\alpha_L^2\pi}{m\kappa}-\frac{3\hbar^2\alpha_L^4 g_{N+k}\pi}{4m^2\kappa^4\dot{x}_k^+}}\right).
\end{eqnarray}
By numerically simulating these motion processes, we can get values of $W_{N+k^+}$, $W_{N+k}^-$ and $E_\gamma$.
By testing whether Eq. (\ref{eq13}) is satisfied, we can get $A_{min}$, $A_{max}$, and then $\overline{A}$ of stable self-sustained oscillation.

\bibliography{ref}

\end{document}